# EPIC Results from ALICE

John W. Harris[a] for the ALICE Collaboration

[a]*Physics Department, Yale University, New Haven CT 06520-8120 U.S.A.*

**Abstract.** An overview is presented of the recent heavy ion results from the ALICE experiment at the Large Hadron Collider. These new results are placed in perspective with those from the Relativistic Heavy Ion Collider experiments.

## INTRODUCTION

The first heavy ion run at the LHC took place in November 2010. Lead (Pb) beams were collided at $\sqrt{s_{NN}}$ = 2.76 TeV with increasing luminosities throughout the run, reaching approximately $2 \times 10^{25}$cm$^{-2}$s$^{-1}$ at the end of the run. ALICE recorded approximately 30M (million) nuclear minimum bias (MB) Pb-Pb interactions to tape. In 2010, ALICE also collected 800M MB, 100M muon-triggered events and 20M high multiplicity pp interactions at $\sqrt{s}$ = 7 TeV, and 8M MB pp interactions at 0.9 TeV. In early 2011, a short pp run was taken at the heavy ion energy of 2.76 TeV in which ALICE accumulated 70M MB events with an integral luminosity of 20 nb$^{-1}$ for rare triggers, providing a pp reference data set for the Pb-Pb data. Results from the pp runs are summarized elsewhere.[1] The remainder of this document will provide a condensed overview of the ALICE heavy ion results to date.

## ALICE EXPERIMENT

The ALICE experiment [2] at the LHC was designed and optimized for heavy ion physics measurements with the large particle multiplicities anticipated with ions at LHC energies. The layout of the ALICE experiment is shown in Fig. 1.

At the heart of ALICE are comprehensive tracking and particle identification systems covering mid-rapidity ($|\eta| \leq 0.9$) with complete azimuthal coverage ($\Delta\phi = 2\pi$) inside the large L3 solenoidal magnet (B = 0.5 T). These central detectors facilitate identification and momentum measurement of hadrons, electrons and photons. The Inner Tracking System (ITS) is closest to the interaction vertex and consists of 6 layers of silicon tracking detectors - two barrels each of silicon pixel detectors, silicon drift detectors and silicon micro-strip detectors. A large time projection chamber (TPC) surrounds the ITS, followed by transition radiation detectors (TRD) and a time-of-flight (TOF) system. The forward muon spectrometer is located at $2.4 \leq \eta \leq 4.0$ inside a 3 T-m dipole magnet for measurements of open heavy flavor and quarkonia.

There are specialized detectors in ALICE with a variety of coverage. The high resolution photon spectrometer (PHOS), covering $|\eta| \leq 0.12$ and $\Delta\phi = 60°$, and the electromagnetic calorimeter (EMCal), installed in early 2011 covering $|\eta| \leq 0.7$ and $\Delta\phi = 100°$, are designed to measure photons, $\pi^0$ and $\eta$ mesons, and jets (in the EMCal). The HMPID, a system of ring imaging Cherenkov detectors, extends the particle ($\pi$, K, p) identification for momenta beyond that of the time-of-flight detector system at mid-rapidity. There are also several smaller detector systems (PMD, FMD, ZDC, T0, V0) in the forward directions, for triggering and multiplicity measurements.

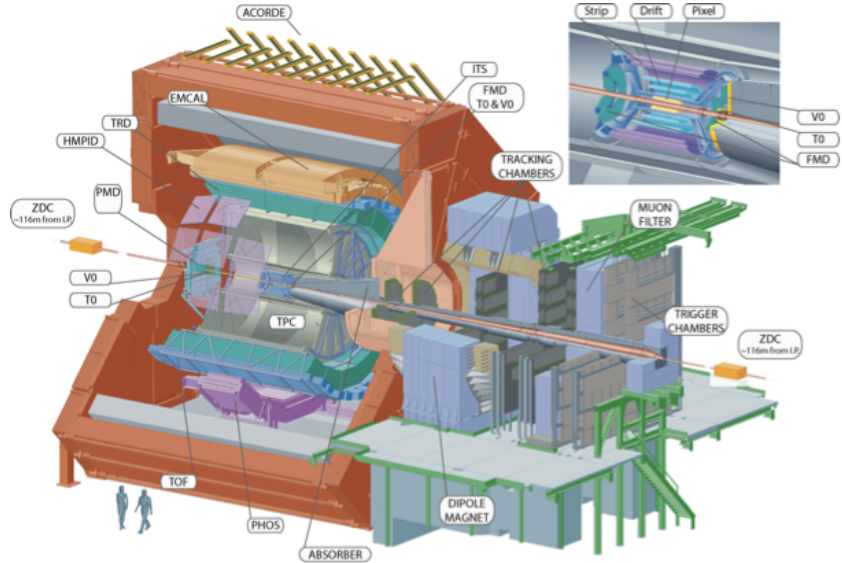

**FIGURE 1**. Schematic layout of the ALICE experiment. See text for description.

# GLOBAL OBSERVABLES

## Charged Particle Multiplicities

The pseudo-rapidity density of produced charged particles places constraints on particle production mechanisms and provides an estimate of the initial energy density. Its dependence on the center-of-mass energy and the size of the system depends both on hard scattering and soft processes, as well as their interplay. Displayed in Fig. 2 (left panel) is the measured charged particle multiplicity per participant pair for central (0-5%) collisions as a function of $\sqrt{s_{NN}}$, plotted over two decades of energy. The data, described in the legend, and corresponding fit curves are presented for AA and pp collisions.[3] The Pb-Pb multiplicity at $\sqrt{s_{NN}}$ = 2.76 TeV from the LHC exhibits an increase by a factor of 1.9 above the pp curve at the same energy and by a factor of 2.2 above the values obtained at 0.2 TeV at RHIC. An estimate of the Pb-Pb energy density $\varepsilon$ for formation time $\tau$ assuming Bjorken expansion indicates $\varepsilon\tau \sim 16$ GeV/(fm$^2$c) at the LHC energy or approximately an increase of 2.5 above that at RHIC.

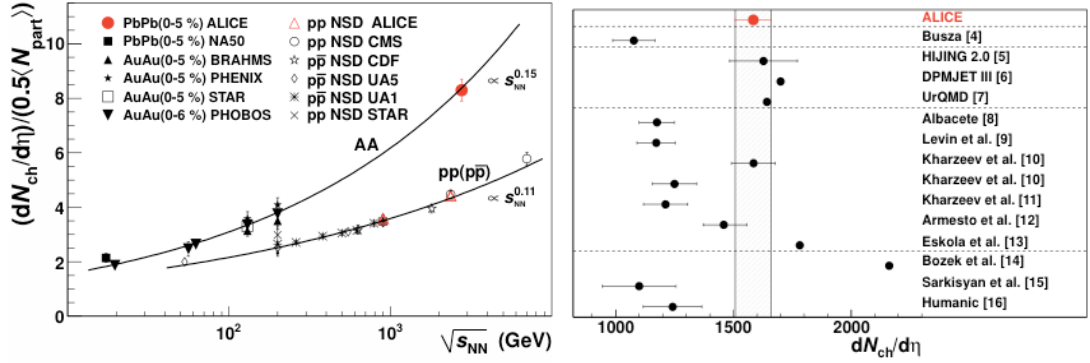

**FIGURE 2.** Left panel: Charged particle pseudorapidity density per participant pair for central AA and non-singly diffractive pp collisions as a function of $\sqrt{s_{NN}}$. Curves are fits to the two data sets. Right panel: Comparison of $dN_{ch}/d\eta$ for ALICE Pb-Pb measurement at top with model predictions grouped below by similar theoretical approaches separated by dashed lines. See text and Ref. [3] for details and model references.

The ALICE result for $dN_{ch}/d\eta$ at midrapidity for Pb–Pb at $\sqrt{s_{NN}} = 2.76$ TeV is $dN_{ch}/d\eta = 1584 \pm 4$ (stat.) $\pm 76$ (sys.). This is shown in Fig. 2 (right panel) with predictions from various models. As a whole the perturbative QCD-inspired Monte Carlo models (figure, notation and references used in Fig. 2 are from Ref. [3]) based either on HIJING, the Dual Parton Model, or Ultrarelativistic Quantum Molecular Dynamics are consistent with the ALICE data.

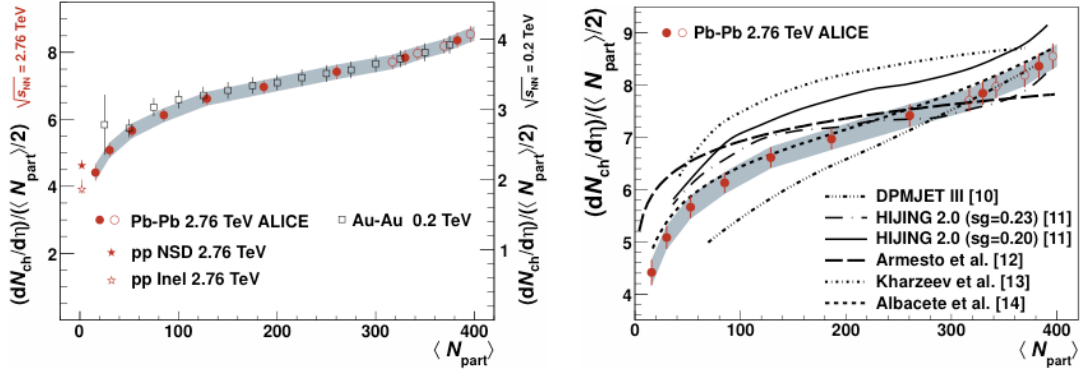

**FIGURE 3.** Left panel: Charged particle pseudorapidity density per participant pair for Pb–Pb and pp collisions at $\sqrt{s_{NN}} = 2.76$ TeV (left vertical scale) and Au–Au collisions at 0.2 TeV (right vertical scale), plotted as a function of $\langle N_{part} \rangle$. Statistical errors are negligible, uncorrelated uncertainties indicated by error bars, and correlated uncertainties as gray band. Right panel: Comparison of model predictions for Pb–Pb at $\sqrt{s_{NN}} = 2.76$ TeV with the ALICE data from left panel. Note offset zero. See Ref. [4].

Displayed in Fig. 3 (left panel) are the $(dN_{ch}/d\eta)/(\langle N_{part} \rangle/2)$ for Pb-Pb collisions at $\sqrt{s_{NN}} = 2.76$ TeV (refer to scale on left ordinate) and Au–Au collisions at 0.2 TeV (right ordinate scale) as a function of $\langle N_{part} \rangle$, i.e. centrality. The centrality dependence is strikingly similar for the ALICE and RHIC data. A comparison of these data to model predictions can be seen in Fig. 3 (right panel). Both the two-component HIJING 2.0 model with strong impact parameter dependent gluon shadowing and the "Albacete" model with a color glass condensate reasonably describe the data. A calculation based on the two-component Dual Parton Model (DPMJET III), with string fusion, rises more strongly with centrality than observed. The remaining models, all different implementations of the saturation picture, show a characteristically weak dependence of multiplicity on centrality. For more details see Refs. [4] and [5].

## Identified Particle Ratios

The ratios of the multiplicities of particles of different species created in Pb-Pb collisions at the LHC can provide information on the degree of thermalization and the chemical equilibrium values in these collisions. A priori, differences are not expected if particle production is dominated by production at chemical freezeout. ALICE has measured the $K^-/\pi^-$ and $p^-/\pi^-$ ratios as a function of $dN_{ch}/d\eta$ for Pb-Pb at $\sqrt{s_{NN}}$ = 2.76 TeV and found the same values and dependence for $dN_{ch}/d\eta > 80$ as in Au–Au collisions at 0.2 TeV.[6] Additional multiplicity ratios are expected soon.

## Identified Particle Spectra

The transverse momentum spectra of identified pions, kaons and protons were measured for both charge states (positive and negative) in Pb-Pb collisions at $\sqrt{s_{NN}}$ = 2.76 TeV in ALICE. Results are presented in Fig. 4 (left panel) for $\pi^-$, $K^-$, $p^-$ and $K^0_s$ in 0-5% central collisions. Results from STAR and PHENIX are also shown for Au–Au collisions at 0.2 TeV. The ALICE data exhibit a stronger power law dependence, as expected, especially for anti-protons compared to RHIC. This suggests stronger radial flow at the LHC. Blast wave fits to spectra indicate an increase of the average radial boost velocity up to (2/3)c and a decrease in the kinetic freezeout temperature to just below 100 MeV relative to RHIC data as seen in Fig. 4 (right panel).

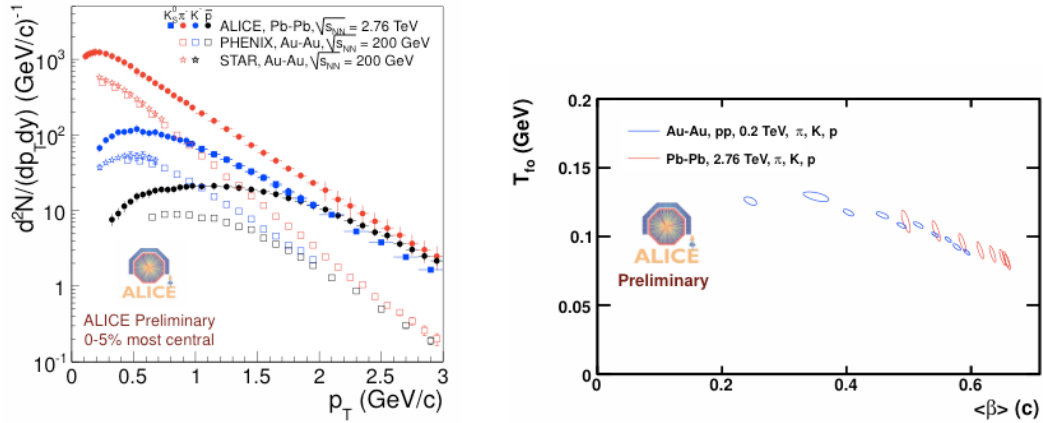

**FIGURE 4.** Left panel: Transverse momentum spectra of various identified particles in ALICE and at RHIC as described in the legend. Right panel: 1 σ-contours for best-fit values for the kinetic freezeout temperature and the average radial boost velocity from the Blast Wave model.[7]

It is of interest to investigate whether the "baryon anomaly" observed at RHIC is present at the LHC. The observation at RHIC of enhanced baryon to meson ratios for transverse momenta up to about 7 GeV/c has been described in terms of quark recombination. The $\Lambda/ K^0_s$ ratios measured in ALICE as a function of $p_T$ are shown in

Fig. 5 (left panel) for different centralities in Pb–Pb collisions at $\sqrt{s_{NN}}$ = 2.76 TeV and for pp at $\sqrt{s}$ = 0.9 and 7 TeV. The $\Lambda$/ $K^0_s$ ratio in peripheral Pb–Pb collisions is slightly larger than that for pp interactions at $\sqrt{s}$ = 7 TeV where $\Lambda$/ $K^0_s$ ~ 0.5. For more central collisions, the $\Lambda$/ $K^0_s$ ratio increases and develops a maximum, reaching a ratio $\Lambda$/ $K^0_s$ ~ 1.5 for $p_T$ ~ 3-3.5 GeV/c in 0-5% central collisions. A comparison with results[a] from RHIC for 0-5% central and 60-80% peripheral Au-Au collisions in Fig. 5 (right panel) shows only slightly larger ratios at the LHC, but perhaps a persistence of ratios larger than those of pp out to higher $p_T$.

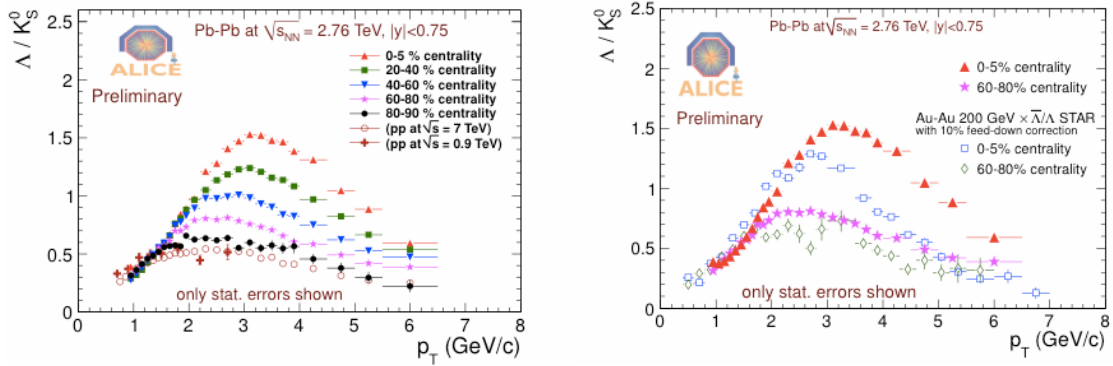

**FIGURE 5.** Left panel: $\Lambda$/ $K^0_s$ ratios at midrapidity as a function of transverse momentum for various centralities in Pb-Pb collisions at $\sqrt{s_{NN}}$ = 2.76 TeV. Ratios are also presented for minimum bias pp events at 0.9 and 7 TeV. Right panel: Comparison of central and peripheral collision ratios from the left panel with ratios in similar Au-Au collisions at $\sqrt{s_{NN}}$ = 0.2 TeV. See text for details.

# COLLECTIVE FLOW

## Charged Particle Elliptic Flow

Elliptic flow ($v_2$) measurements at RHIC indicate that multiple interactions within a very short timescale create a strongly-interacting medium of low viscosity in these collisions, more precisely a low value of the ratio shear viscosity ($\eta$) / entropy (s). Furthermore, since the temperature dependence of $\eta/s$ of this medium is unknown, a measurement of the elliptic flow at the LHC and determination of $\eta/s$ are needed. In Fig. 6 (left panel) is the "world's data" on the elliptic flow $v_2$ integrated over $p_T$ as a function of $\sqrt{s_{NN}}$.[8] The integrated elliptic flow of charged particles at the LHC increases by ~ 30% over that of the top energy at RHIC. Thus, the hot medium created in Pb-Pb collisions at the LHC behaves very much like that at RHIC and should provide constraints on the temperature dependence of $\eta/s$.

Differential elliptic flow measurements are sensitive to the dynamical evolution and freezeout conditions of the system. Displayed in Fig. 6 (right panel) is the elliptic flow $v_2$(4) determined from the 4-particle cumulant as a function of $p_T$ for ALICE data [8] at $\sqrt{s_{NN}}$ = 2.76 TeV and STAR data at $\sqrt{s_{NN}}$ = 200 GeV, 62.4 GeV and 39 GeV [9]. The $p_T$ dependence of $v_2$(4) appears essentially identical for 20-30% centrality Pb-Pb collisions at the LHC and Au-Au collisions at RHIC from $\sqrt{s_{NN}}$ = 2.76 TeV down to

---

[a] STAR data are multiplied by 0.8 to account for the anti-baryon/baryon ratio and a 10 % feed-down correction is made.

39 GeV. That the differential elliptic flow would remain the same over two orders of magnitude in energy was not anticipated. The increase in the integrated flow observed in Fig. 6 (left panel) therefore must be due to an increase in the average transverse momentum, partly attributed to increased radial flow, as seen in the spectra of Fig. 4 (left panel).

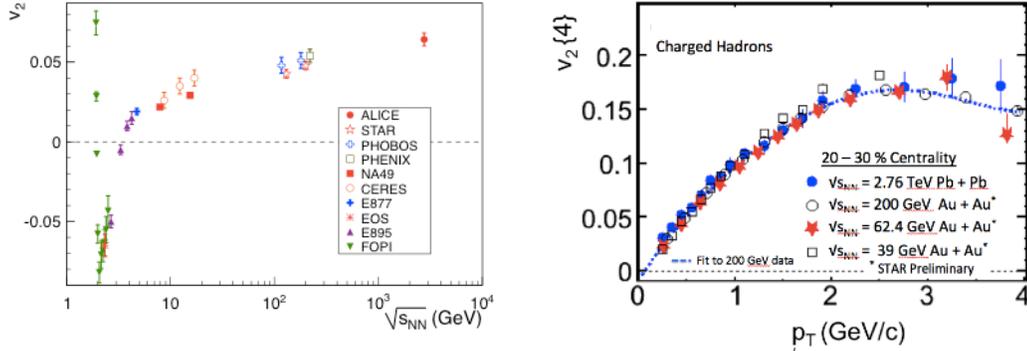

**FIGURE 6.** Left panel: Integrated elliptic flow ($v_2$) as a function of $\sqrt{s_{NN}}$ for central heavy ion collisions (from Ref. [8]). Right panel: The elliptic flow determined from the 4-particle cumulant as a function of $p_T$ for 20-30% centrality ALICE data [8] at $\sqrt{s_{NN}} = 2.76$ TeV and STAR data [9] at $\sqrt{s_{NN}} = 39$, 62.4, and 200 GeV.

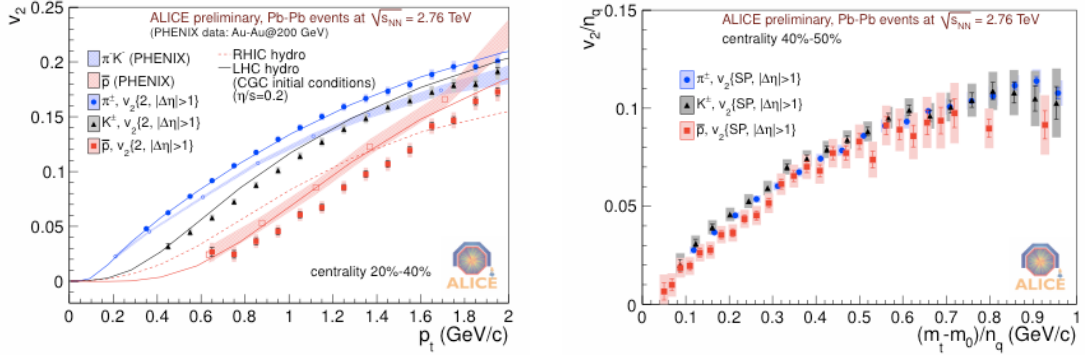

**FIGURE 7.** Left panel: Elliptic flow ($v_2$) as a function of transverse momentum for Pb-Pb collisions at $\sqrt{s_{NN}} = 2.76$ TeV from ALICE and Au-Au collisions at $\sqrt{s_{NN}} = 200$ GeV in PHENIX for 20 – 40% centrality. Also shown are results from hydrodynamics.[10] Right panel: $v_2$ per quark as a function of transverse kinetic energy per quark as measured by $v_2$ from $\pi^\pm$, $K^\pm$, and p̄ in 40 – 50 % centrality from ALICE.

## Identified Particle Elliptic Flow

The observed increase in radial flow at the LHC also results in a stronger particle mass dependence of the elliptic flow at the LHC than at RHIC. This can be observed in Fig. 7 (left panel), where the $v_2(p_T)$ is presented for pions, kaons and anti-protons from ALICE and from PHENIX. The "mass splitting," or the separation between the differential elliptic flow for different masses is observed to be larger at the LHC than at RHIC and also than that predicted by hydrodynamics.[10] Furthermore, the "quark number scaling" of $v_2(p_T)$ observed at RHIC appears to be broken at the LHC as seen by the variation of the three curves in Fig. 7 (right panel). At $p_T > 2$ GeV/c, the quark number scaling of $v_2(p_T)$ still holds at the LHC (not shown) [11], supporting a description in terms of coalescence.

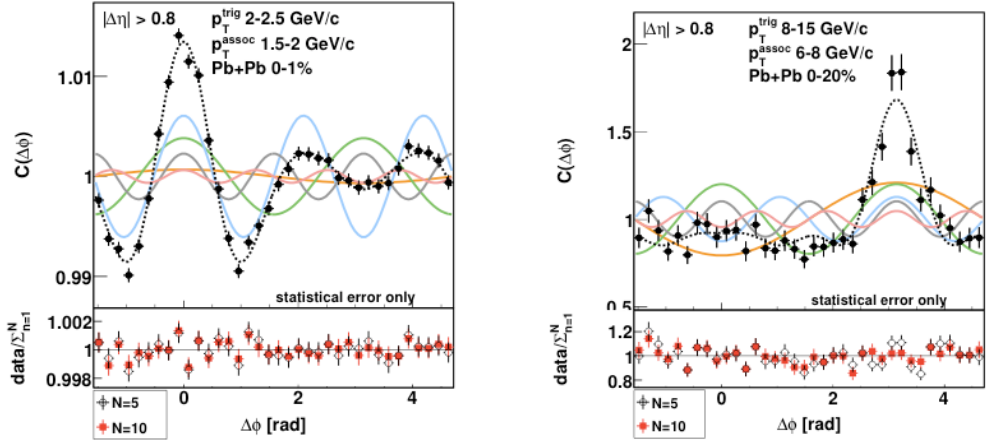

**FIGURE 8.** Azimuthal projections of correlation function data over $0.8 < |\Delta\eta| < 1.8$. Left panel: low $p_T$ triggers (see legend). Right panel: high $p_T$ triggers. Dashed curve is result of a Fourier decomposition (fit) using the first 5 components (color curves). At bottom of figures are ratios of data to fit for cases of n = 5 and 10, with little observable difference (note more sensitive scale on left compared to right).

## Fourier Decomposition of Azimuthal Di-hadron Correlations

Triggered di-hadron correlations can be used to identify and separate the regions of phase space in which collective effects dominate from those where jet-related correlations are dominant. If the two-particle correlation is connected by a common plane of symmetry then the factorization relation, $V_{n\Delta}(p_{Tassoc}, p_{Ttrig}) = v_n(p_{Tassoc}) \cdot v_n(p_{Ttrig})$ should be satisfied, thus signifying the dominance of collective effects. This relation is not expected to hold for jet-related correlations such as those from jet fragmentation. Displayed in Fig. 8 are the correlation functions $C(\Delta\phi) = dN_{assoc}/(N_{pairs} \cdot d(\Delta\phi))$ for individual events after normalizing by the same correlation in mixed events and integration over $0.8 < |\Delta\eta| < 1.8$ for low and high $p_T$ trigger conditions. A Fourier decomposition of these long-range (in $\eta$) correlations using n = 10 or even 5 harmonics yields a good fit (dashed curve) to the low $p_T$ trigger data as seen in Fig. 8 (left panel), but not for the high $p_T$ trigger in Fig. 8 (right panel) where the recoil jet dominates the correlation. Note also that the data in Fig. 8 (left panel) correspond to the most central data set (0 − 1 % centrality) and exhibit a double-hump structure on the away-side of the trigger particle. This complicated structure appears to be due simply to contributions of the various harmonic components, most probably fluctuations in the initial state configuration. Furthermore, the factorization hypothesis works well for $p_T$ up to ∼ 3 − 4 GeV/c, but fails at higher $p_T$ where jet-like effects dominate correlations. For a more detailed analysis and discussion, see Refs. [12],[13].

## SPACE-TIME EVOLUTION OF THE SYSTEM

The hydrodynamic approach describes reasonably well the momentum distribution of particles in RHIC and LHC heavy ion collisions. However, their spatial distribution at the time of decoupling is more difficult to assess, as it is determined by the initial temperature and equation of state of the system. The expansion rate and the spatial extent at decoupling can be accessed in experiment using intensity interferometry, or Bose–Einstein correlations of identical bosons emitted close in phase space. A homogeneity volume V can be measured from the 3-dimensional spatial extent of the system from such correlation analyses. ALICE finds this volume to be V ~ 300 fm$^3$ for central Pb-Pb collisions at $\sqrt{s_{NN}}$ = 2.76 TeV.[14] This is about twice that measured at RHIC in central Au-Au collisions at $\sqrt{s_{NN}}$ = 200 GeV. In the same ALICE analysis, the decoupling time is found to be 10 − 11 fm/c for central Pb-Pb collisions, or about 40 percent longer than at RHIC. Thus, the fireball formed in heavy ion collisions at the LHC is hotter (higher ε), lives longer and expands to a larger size at freeze-out than at RHIC.

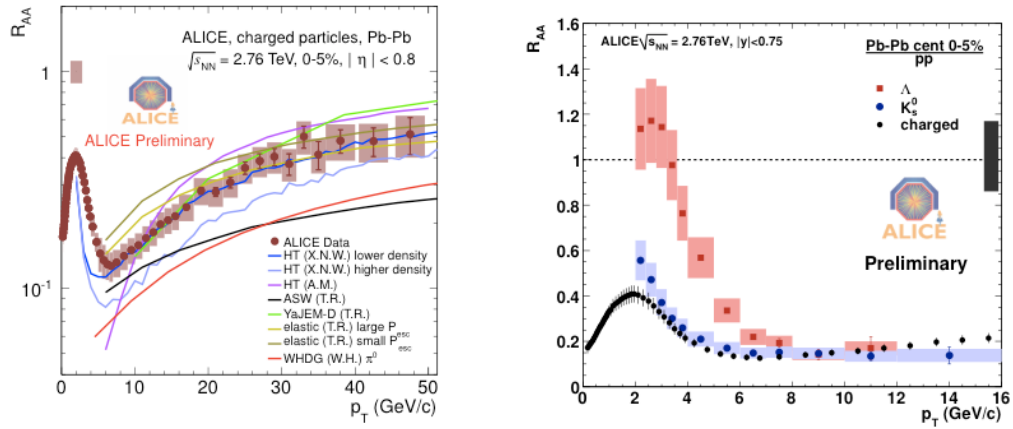

**FIGURE 9.** Nuclear modification factor ($R_{AA}$) as a function of transverse momentum for 0 − 5% central Pb-Pb collisions at $\sqrt{s_{NN}}$ = 2.76 TeV. Left panel: Charged particles. Right panel: Λ and $K_s^0$.

# HARD PROBES

## Charged and Identified Particle Nuclear Modification Factors

Charged particles in central Au-Au collisions from RHIC at $\sqrt{s_{NN}}$ = 200 GeV are found to be suppressed by a factor of 4 − 5 for transverse momenta ($p_T$) above a few GeV/c compared to a superposition of independent proton-proton collisions. Since parton fragmentation is the dominant production mechanism of high $p_T$ particles, this suggests strong parton energy loss in the medium. The nuclear modification factor $R_{AA}(p_T)$ for central Pb+Pb collisions at $\sqrt{s_{NN}}$ = 2.76 TeV has been measured by ALICE and is presented in Fig. 9 (left panel). See Ref. [12] for definition of $R_{AA}$ and details. The suppression at the LHC as measured by ALICE is slightly larger than at RHIC, with a sharp rise for $p_T$ > 6 GeV/c. Also shown are results of calculations from energy loss models (see Ref. [12]) that have been constrained to fit the RHIC $R_{AA}$ data. These models reproduce the rising trend in the data, but the absolute values vary depending upon the degree of quenching and the quenching mechanisms.

Measurements of the nuclear modification factor for identified particles are underway in ALICE. Shown in Fig. 9 (right panel) are the $R_{AA}(p_T)$ for Λ and $K_s^0$ in 0-

5% central Pb+Pb collisions at $\sqrt{s_{NN}}$ = 2.76 TeV compared to charged particles. A distinct difference between the baryon ($\Lambda$) and meson ($K_s^0$) are observed. Baryons are not as suppressed as mesons at low to intermediate $p_T$. This is expected from the stronger radial boost from flow for baryons and the resultant "baryon anomaly" observed in Figs. 4 and 5, respectively. Similar results have been seen at RHIC.

## Heavy Quark Nuclear Modification Factors

Particles containing quarks of heavy flavor (most prominently c and b quarks at the LHC) are expected to probe the hot, dense phase of ultra-relativistic heavy ion collisions. Leading particles with open charm or beauty should be sensitive to properties of the medium, such as the energy density, and provide information on the interaction of partons with the medium via in-medium energy loss. Quarkonia are expected to be sensitive probes of the initial temperatures and deconfinement via color screening effects.

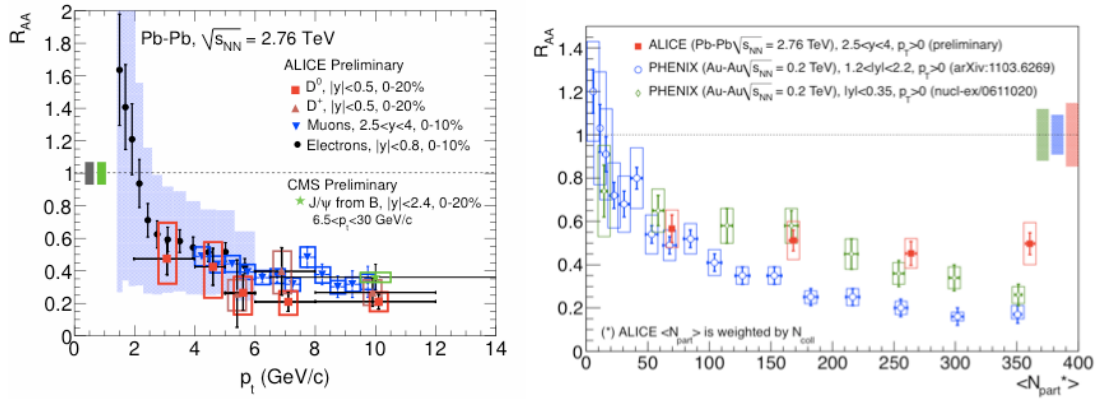

**FIGURE 10.** Left panel: Nuclear modification factor ($R_{AA}$) for particles associated with heavy flavor production and decays as described in legend as a function of transverse momentum for central Pb-Pb collisions at $\sqrt{s_{NN}}$ = 2.76 TeV in ALICE and for J/$\psi$ from B-decays in CMS.[15] Right panel: $R_{AA}$ as a function of $\langle N_{part} \rangle$ for J/$\psi$ in Pb-Pb collisions in ALICE at $\sqrt{s_{NN}}$ = 2.76 TeV and Au-Au collisions in PHENIX[16] at $\sqrt{s_{NN}}$ = 0.2 TeV.

ALICE has measured the nuclear modification factors for hadrons with heavy flavor in Pb-Pb central collisions at $\sqrt{s_{NN}}$ = 2.76 TeV. $R_{AA}(p_T)$ are presented in Fig. 10 (left panel) for D-mesons at mid-rapidity (utilizing displaced decay vertex reconstruction) and for electrons and muons at mid- and forward-rapidity, respectively. D-mesons are as suppressed as charged particles (mostly pions), as seen at RHIC. Prompt J/$\psi$ from B-decays as measured in CMS[15] are similarly suppressed, as also shown in Fig. 10 (left panel). Displayed in Fig. 10 (right panel) are $R_{AA}(p_T)$ as a function of centrality ($\langle N_{part} \rangle$) for J/$\psi$ from ALICE in Pb-Pb collisions at $\sqrt{s_{NN}}$ = 2.76 TeV and from PHENIX[16] in Au-Au collisions at $\sqrt{s_{NN}}$ = 0.2 TeV. A weaker dependence with centrality is observed in ALICE at the LHC than observed in PHENIX at RHIC. In the most central collisions (large $\langle N_{part} \rangle$) the $R_{AA}(p_T)$ is a factor of 2 larger at the LHC than at RHIC for forward-rapidities. These differences persist at midrapidity, but to a lesser degree.

# CONCLUSIONS

ALICE has found that central collisions of Pb-Pb at the LHC produce a charge multiplicity density per participant pair that is about 2.2 times larger and an energy density about 3 times higher than at RHIC. The RHIC Baryon Anomaly still exists and appears similar at LHC and RHIC. Stronger radial flow is seen at LHC than at RHIC.

The integrated elliptic flow ($v_2$) increases from RHIC to the LHC, however the centrality and $p_T$ dependence of $v_2$ is the same for Pb-Pb at $\sqrt{s_{NN}} = 2.76$ TeV at the LHC and Au-Au at RHIC energies down to $\sqrt{s_{NN}} = 39$ GeV. There is larger differential $v_2(p_T)$ mass splitting, especially for (anti-)protons, at the LHC with $v_2(p^-) > v_2(\pi^-)$ up to $p_T \sim 8$ GeV/c. The quark scaling of $v_2$ observed at RHIC does not work at $p_T < 2$ GeV/c at the LHC. The $v_2(p_T)$ can be described by viscous hydrodynamics with $\eta/s \sim 0.2$. A Fourier decomposition of long-range correlations accounts for low-$p_T$ triggered azimuthal angular correlations, but not for high-$p_T$ ($p_T > 3$-4 GeV/c) triggers due to the recoil jet, signifying the significant role of initial state fluctuations in the low $p_T$ data.

Central Pb-Pb collisions at LHC have a 2 times larger volume ($\sim 300$ fm$^3$) and a 1.4 times longer lifetime ($\sim 10$-11 fm/c) compared to central Au-Au collisions at RHIC.

Central collisions of Pb-Pb at the LHC exhibit strong suppression of charged particles ($R_{AA} \sim 0.1$-0.2), D-mesons ($R_{AA} \sim 0.2$-0.3), B-mesons ($R_{AA} \sim 0.3$-0.4), and J/$\psi$ ($R_{AA} \sim 0.5$). The J/$\psi$ is suppressed although less than at RHIC. These results provide strong evidence for parton energy loss and their relative differences present interesting and difficult challenges to theory.

# AKNOWLEDGMENTS